\documentclass[conference]{IEEEtran}
\ifCLASSINFOpdf
   \usepackage[pdftex]{graphicx}
   \graphicspath{{../pdf/}{../jpeg/}}
   \DeclareGraphicsExtensions{.pdf,.jpeg,.png}
\else
   \usepackage[dvips]{graphicx}
   \graphicspath{{../eps/}}
   \DeclareGraphicsExtensions{.eps}
\fi
\usepackage{fixltx2e}
\usepackage{amsfonts}
\usepackage{amsmath,amssymb,amsfonts}
\usepackage{cite} 
\usepackage{graphicx}  
\usepackage{subfigure} 
\usepackage{amsmath,amssymb,amsfonts}
\usepackage{algorithmic}
\usepackage{textcomp}
\usepackage{algorithm}
\usepackage{multirow}
\usepackage{algorithmic}
\usepackage{bm}
\usepackage{cite}
\usepackage{float}
\usepackage{url}
\usepackage[pdftex]{graphicx}
\usepackage{diagbox}
\usepackage{graphics}
\usepackage{setspace}
\usepackage{caption}
\usepackage{threeparttable}
\usepackage[T1]{fontenc}
\usepackage{aecompl}

\usepackage{fancyhdr}

\begin{document}
%

\title{Affine-Doppler Division Multiplexing for High-Mobility Wireless Communications Systems}



\author{\IEEEauthorblockN{Yuanfang Ma, Zulin Wang, Peng Yuan, Qin Huang and Yuanhan Ni\IEEEauthorrefmark{1}}
	\IEEEauthorblockA{School of Electronic and Information Engineering, Beihang University, Beijing, 100191, China\\
		Email: yuanfangma@buaa.edu.cn, wzulin@buaa.edu.cn, yuanpeng9208@buaa.edu.cn, \\
		qhuang.smash@gmail.com, yuanhanni@buaa.edu.cn (corresponding author) \\
}}

\maketitle

\thispagestyle{fancy}
\fancyhead{}
\lhead{}
\cfoot{}
\rfoot{}

\begin{abstract}
Affine Frequency Division Multiplexing (AFDM) has been regarded as a candidate integrated sensing and communications (ISAC) waveform owing to its superior communication performance, outperforming the Orthogonal Time-Frequency Space (OTFS) that has been researched for a longer time. However, since the above two waveforms are incompatible with each other, the state-of-the-art methods well-designed for OTFS may not be directly applicable to AFDM. This paper introduces a new orthogonal multicarrier waveform, namely Affine-Doppler Division Multiplexing (ADDM), which can provide a generic framework and subsume the existing OTFS and AFDM as a particular case. ADDM modulating information symbols in the Affine-Doppler (A-D) domain based on a two-dimensional (2D) transform can enjoy both excellent unambiguous Doppler and Doppler resolution, which is the same as AFDM but outperforms OTFS. Moreover, benefiting from the 2D transform, the symbols block of ADDM in the A-D domain undergoes a 2D cyclic shift produced by the delay and the Doppler of the channel, similar to the 2D cyclic shift in the delay-Doppler domain of cyclic prefix (CP)-OTFS. This offers a potential to directly apply the state-of-the-art methods well-designed for OTFS and AFDM to ADDM. Numerical results show that ADDM achieves comparable BER performance with AFDM but outperforms OTFS in high-mobility scenarios.
\end{abstract}


%
\IEEEpeerreviewmaketitle

\section{Introduction}


Next-generation wireless communications systems (beyond 5G/6G) are expected to significantly improve spectral and energy efficiencies, support ubiquitous connections of everything and maintain reliable communications in high-mobility scenarios\cite{IMT2021White}. The integrated sensing and communications (ISAC) technique is one of the critical enablers of 5G/6G due to its ability to improve spectral and energy efficiencies and obtain information about the environment \cite{liu2020joint,SLiu2025Random}.

ISAC waveform design plays a vital role in the ISAC system. Multicarrier waveforms have been widely studied as ISAC waveforms due to their advantages of high communication spectral efficiency, robustness against multipath fading, good ambiguity characteristics, etc. For example, the orthogonal frequency division multiplexing (OFDM) waveform has been used to simultaneously realize sensing and communications (S$\&$C) functionalities\cite{sturm2011waveform,zeng2020joint}. However, the communications bit error rate (BER) performance of OFDM waveform may deteriorate severely in high-mobility scenarios\cite{Raviteja2018Interference,bemani2023affine}.

To improve the performance of ISAC systems in high-mobility scenarios, orthogonal time-frequency space (OTFS)-based ISAC waveforms have been investigated  \cite{Raviteja2018Interference,Gaudio2020On}. OTFS waveform spreads information symbols in the delay-Doppler (D-D) domain\cite{hadani2017orthogonal,Raviteja2018Interference}. Consequently, OTFS can deal with significant Doppler shifts and obtain both time and frequency diversities in doubly selective channels. Consequently, OTFS can significantly improve BER performance in doubly selective channels compared with OFDM. Thus, OTFS has been regarded as a potential candidate ISAC waveform, and many methods have been well designed for OTFS to meet the requirements of 5G/6G in terms of channel estimation, symbol detection, coding, multi-access, sensing parameter estimation, etc. However, the unambiguous Doppler of OTFS is limited by the subcarrier spacing. If the difference between Doppler shifts of two paths (with identical delays) is equal to the subcarrier spacing, these two paths will overlap in the D-D domain, which degrades BER performance. While reducing the number of subcarriers can increase the subcarrier spacing and thereby increase the unambiguous Doppler, it will reduce the communication spectral efficiency due to the existence of a cyclic prefix (CP). Recently, frequency Doppler division multiplexing (FDDM) has been proposed that modulates symbols in the frequency-Doppler (F-D) domain \cite{gong2024frequency}, whose unambiguous Doppler can break through the subcarrier spacing. However, FDDM can only achieve partial diversity in frequency-selective channels, similar to OFDM.

A chirp multicarrier waveform, namely affine frequency division multiplexing (AFDM), has been proposed by multiplexing information symbols in the affine domain based on discrete affine Fourier transform (DAFT) \cite{bemani2023affine,luo2024afdm,YTao2025IdxAFDM,Hyin2024MIMOAFDM}. By setting appropriate parameters, the unambiguous Doppler of AFDM can be several times the subcarrier spacing. Moreover, AFDM can achieve full diversity in the doubly selective channels with distinct delays or integer Doppler shifts (i.e., the integer part of Doppler shifts normalized by subcarrier spacing). Compared with OTFS, AFDM has higher spectral efficiency due to less channel pilot overhead. AFDM-ISAC systems have been investigated in \cite{ni2022afdm,bemani2024integrated}, showing excellent sensing performances even in high mobility scenarios. However, since AFDM is not compatible with OTFS, the state-of-the-art methods well-designed for OTFS may not be directly applicable to AFDM.

This paper introduces Affine-Doppler Division Multiplexing (ADDM), a new orthogonal multicarrier waveform, to provide a generic framework for ISAC waveform. Specifically, the information symbols of ADDM are modulated in the Affine-Doppler (A-D) domain based on a two-dimensional (2D) transform, i.e., DAFT and discrete Fourier transform. Due to the existence of DAFT, ADDM can inherit the merits of AFDM and thus enjoy both excellent unambiguous Doppler and Doppler resolution, outperforming OTFS. 
The compatibility analysis of ADDM reveals that the proposed ADDM can subsume existing OFDM, FDDM, OTFS and AFDM as a particular case.
This offers a potential to directly apply the state-of-the-art methods well-designed for OTFS and AFDM to ADDM. Numerical results show that the proposed ADDM achieves comparable BER performance with AFDM but outperforms OTFS in high-mobility scenarios.
\section{Preliminaries}

\subsection{Discrete Affine Fourier Transform}

We briefly review DFT and DAFT, which form the basis of OFDM, OTFS and AFDM.

Discrete Fourier transform (DFT) is defined as 
\begin{equation} 
	p_{\mathrm{FT},m} = \frac{1}{{\sqrt {{N_0}} }}\sum\limits_{n = 0}^{{N_0} - 1} {{e^{ - j\frac{{2\pi }}{{{N_0}}}mn}}} {s_n} ,
	\label{eq: DFT}
\end{equation}
where ${s_n}$ and ${p_{\mathrm{FT},m}}$ denote the symbols in the delay domain and the frequency domain, respectively. $m=0,1,\ldots ,{{N}_{0}}-1$ and $n=0,1,\ldots ,{{N}_{0}}-1$. The inverse DFT (IDFT) is given by 
\begin{equation} 
	{s_n} = \frac{1}{{\sqrt {{N_0}} }}\sum\limits_{m = 0}^{{N_0} - 1} {{e^{j\frac{{2\pi }}{{{N_0}}}mn}}} {p_{\mathrm{FT},m}} .
	\label{eq: IDFT}
\end{equation}
DFT and IDFT can be expressed in matrix forms as \cite{bemani2023affine}
\begin{equation} \label{eq:DFT in matrix form}
	\mathbf{p}_\mathrm{FT}=\mathbf{F}_{N_0}\mathbf{s}, \quad \mathbf{s} = \mathbf{F}_{N_0}^{\rm H}\mathbf{p}_\mathrm{FT},	
\end{equation}
where $\mathbf{p}_\mathrm{FT}=\left(p_{\mathrm{FT},0},p_{\mathrm{FT},1},...,p_{\mathrm{FT},N_0-1} \right)^{\rm T}$ and $\mathbf{s}=\left( {{s}_{0}},{{s}_{1}},...,{{s}_{{{N}_{0}}-1}} \right)^{\rm T}$. ${\bf{F}}_{{N_0}}$ is DFT matrix with entries
\begin{equation} \label{eq: explanation of F in DAFT}
	{{\bf{F}}_{{N_0}}}\left[ {m,n} \right] = \frac{1}{{\sqrt {{N_0}} }}{e^{ - j2\pi \frac{{mn}}{{{N_0}}}}}.
\end{equation}

DAFT is the discretization of affine Fourier transform (AFT), which maps symbols from the delay domain into the affine domain and is defined as \cite[Eq. (12)]{bemani2023affine}
\begin{equation} 
{{p}_{\mathrm{AT},m}}=\frac{1}{\sqrt{{{M}_{0}}}}{{{e}}^{-j2\pi {{c}_{2}}{{m}^{2}}}}\sum\limits_{n=0}^{{{N}_{0}}-1}{{{e}^{-j2\pi (\frac{1}{{{M}_{0}}}mn+{{c}_{1}}{{n}^{2}})}}}{{s}_{n}} ,
	\label{eq: DAFT}
\end{equation}
where ${{p}_{\mathrm{AT},m}}$ is the symbol in the affine domain, $m=0,1,\ldots ,{{M}_{0}}-1$ and $n=0,1,\ldots ,{{N}_{0}}-1$. The inverse DAFT (IDAFT) can be expressed as \cite{bemani2023affine}
\begin{equation} 
{{s}_{n}}=\frac{1}{\sqrt{{{M}_{0}}}}{{{e}}^{j2\pi {{c}_{1}}{{n}^{2}}}}\sum\limits_{m=0}^{{{M}_{0}}-1}{{{e}^{j2\pi (\frac{1}{{{M}_{0}}}mn+{{c}_{2}}{{m}^{2}})}}}{{p}_{\mathrm{AT},m}} .
	\label{eq: IDAFT}
\end{equation}
The periodicity of DAFT is given by \cite{bemani2023affine}
\begin{align} \label{eq: periodicity of Sm}
{{p}_{\mathrm{AT},{m+k{{M}_{0}}}}}&={{e}^{-j2\pi {{c}_{2}}({{k}^{2}}M_{0}^{2}+2k{{M}_{0}}m)}}{{p}_{\mathrm{AT},m}} , \nonumber \\
{{s}_{n+k{{N}_{0}}}}&={{e}^{j2\pi {{c}_{1}}({{k}^{2}}N_{0}^{2}+2k{{N}_{0}}n)}}{{s}_{n}} .
\end{align}
Following \cite{bemani2023affine}, this paper considers ${{M}_{0}}={{N}_{0}}$. At this time, DAFT and IDAFT can be expressed in matrix forms as \cite{bemani2023affine}
\begin{equation} 
\mathbf{p}_{\mathrm{AT}}=\mathbf{As}, \quad \mathbf{s}={{\mathbf{A}}^{\rm H}}\mathbf{p}_{\mathrm{AT}},
	\label{eq: explanation of DAFT in matrix form}
\end{equation}
where $\mathbf{p}_{\mathrm{AT}} =\left( p_{\mathrm{AT},0},p_{\mathrm{AT},1},...,p_{\mathrm{AT},N_0-1} \right)^{\rm T}$, and ${\bf{A}} = {{\bf{\Lambda }}_{{c_2}}}{{\bf{F}}_{{N_0}}}{{\bf{\Lambda }}_{{c_1}}}$ with ${{\mathbf{\Lambda }}_{c}}$ being a diagonal matrix which can be explained as
\begin{equation} 
	{{\mathbf{\Lambda }}_{c}}=\mathrm{diag}\left( {{e}^{-2\pi c{{n}^{2}}}},n=0,1,\ldots ,{N}_{0}-1 \right) .
	\label{eq: explanation of the diagonal in ADFT}
\end{equation}

\subsection{Doubly Selective Channel Model}

In doubly selective channels, information transmission is affected by delay spread and Doppler shift. After transmission over a channel with $P$ paths, the received samples are\cite{bemani2023affine}
\begin{equation} 
\mathbf{r}\left[ n \right]=\sum\limits_{i=1}^{P}{{h}_{i}\mathbf{s}\left[ n-{{l}_{i}} \right]{{e}^{j2\pi {{f}_{i}}n}}}+\mathbf{w}\left[ n \right] ,
	\label{eq: relation between r and s after the channel}
\end{equation}
where ${{h}_{i}}$, ${{l}_{i}}$ and ${{f}_{i}}$ are the complex gain, the delay (in samples) and the Doppler shift (in digital frequencies) of the $i$-th path, respectively. $\mathbf{w} \sim \mathcal{CN}\left( {0,\sigma_n^2} \right)$ is an additive Gaussian noise. 

\section{Affine-Doppler Division Multiplexing}

In this section, we start from analyzing the relation among the A-D, F-D, Time-Frequency (T-F), D-D, Time-Affine (T-A) and Time-Delay (T-D) domains. Based on this, ADDM is introduced, which modulates the information symbols in the A-D domain.

\subsection{Relation Among Different Domains}
\begin{figure}[htbp]
	\centering
	\includegraphics[width=3.3in]{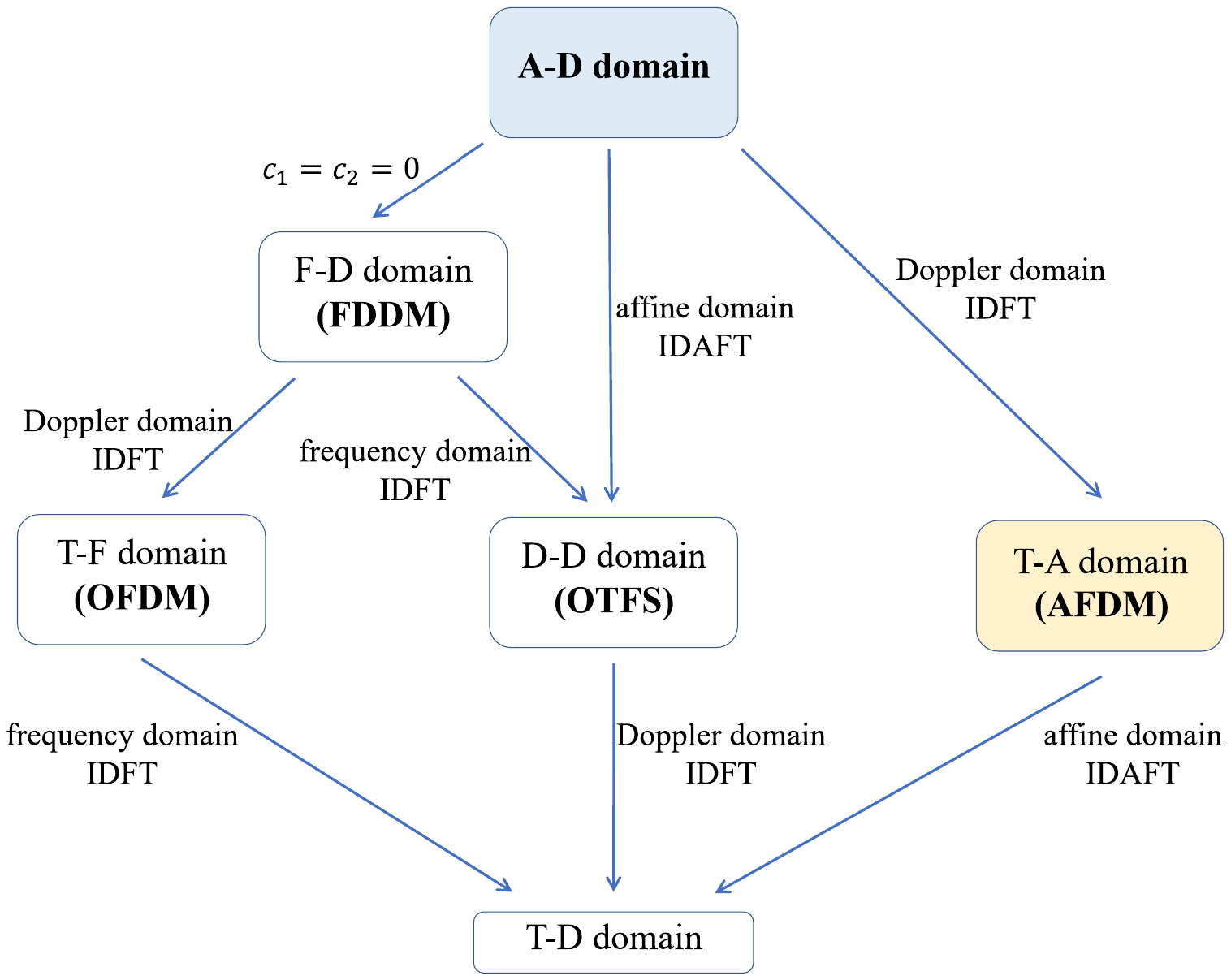}
	\caption{The relation among A-D, F-D, T-F, D-D, T-A and T-D domains.
		\label{fg:relation_diff_domain}}
\end{figure}
The relation among A-D, F-D, T-F, D-D, T-A and T-D domains is shown in Fig. \ref{fg:relation_diff_domain}. 
As we know, OFDM, OTFS, FDDM and AFDM waveforms modulate information symbols in the T-F, D-D, F-D and T-A domains, respectively, which results in different communication performances in doubly selective channels.
We can see that a new domain (namely A-D domain) turns into the F-D domain when $c_1=c_2=0$. Moreover, symbols in the A-D domain can be transformed into the D-D and T-A domains by performing IDAFT in the affine domain and IDFT in the Doppler domain, respectively. Similarly, by performing IDFT in the Doppler domain and IDFT in the frequency domain, symbols in the F-D domain can be transformed into the T-F and D-D domains, respectively. Consequently, the information symbols in the A-D domain can be compatible with information symbols in the F-D, T-F, D-D and T-A domains.   


To the best of our knowledge, there is no waveform that modulates information symbols in the A-D domain. Therefore, this paper introduces a new orthogonal multicarrier waveform, ADDM, to modulate information symbols in the A-D domain. Based on the relation among the A-D, F-D, T-F, D-D and T-A domains analyzed above, the proposed ADDM can potentially subsume existing OFDM, OTFS, FDDM and AFDM as a particular case.



\subsection{The Proposed ADDM Technique}

\begin{figure}[htbp]
	\centering
	\includegraphics[width=3.5in]{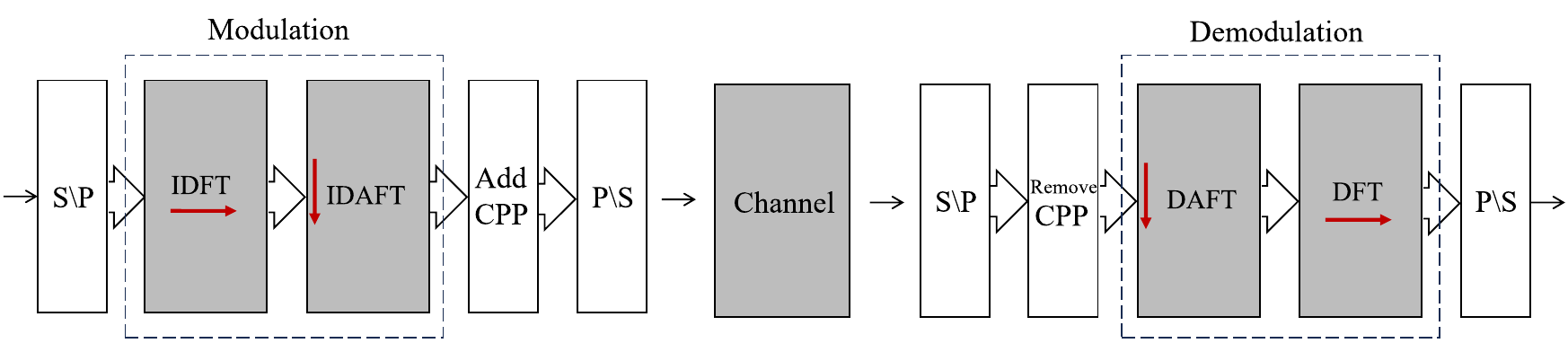}

	\caption{ADDM block diagram.
		\label{fg:ADDM}}
\end{figure}

In this subsection, we propose ADDM waveform and corresponding transceiver. In the ADDM scheme, information symbols are arranged on the A-D domain. The 2D transform, i.e., IDAFT and IDFT, is used to map the information symbols into the T-D domain at the transmitter, while DAFT and DFT are performed at the receiver to obtain the received symbol in the A-D domain. The block diagram of ADDM is shown in Fig. \ref{fg:ADDM}.

\vspace{0.5ex}
\noindent\emph{(1) Modulation}
\vspace{0.5ex}

Consider an information symbol block $\mathbf{X}\left[ n,m \right]$, $n=0,1,\ldots ,N-1,m=0,1,\ldots ,M-1$. The symbols are from a modulation alphabet $\mathbb{A}=\left\{ {{a}_{1}},\ldots ,{{a}_{\left| \mathbb{A} \right|}} \right\}$ (e.g. QAM), which is arranged on the A-D domain. Firstly, IDFT is performed for each row of ${\bf{X}}$, and the resulting matrix in the T-A domain can be obtained as
\begin{equation} 
{\bf{P}}\left[ {m,k} \right] = \frac{1}{{\sqrt M }}\sum\limits_{p = 0}^{M - 1} {{\bf{X}}\left[ {m,p} \right]{e^{j2\pi \frac{{kp}}{M}}}},
\label{eq: relation between S and X in elements}
\end{equation}
where $m = 0,1, \ldots ,N - 1$ and $k = 0,1, \ldots ,M - 1$. Eq. (\ref{eq: relation between S and X in elements}) is rewritten in matrix form as ${\bf{P}} = {\bf{XF}}_M^\mathrm{H}$. 
Secondly, IDAFT is used for each column of $\mathbf{P}$, and the resulting matrix in the T-D domain is given by
\begin{equation} \label{eq: explanation of F1}
	\mathbf{S}\left[ n,k \right]=\frac{1}{\sqrt{N}}\sum\limits_{m=0}^{N-1}{\mathbf{P}}\left[ m,k \right]{{e}^{j2\pi \left( {{c}_{1}}{{n}^{2}}+{{c}_{2}}{{m}^{2}}+\frac{mn}{N} \right)}} ,
\end{equation}
with $n = 0,1, \ldots ,N - 1$ and $k = 0,1, \ldots ,M - 1$. In matrix form, Eq. (\ref{eq: explanation of F1}) is expressed as $\mathbf{S}={{\mathbf{A}}^{\rm H}}\mathbf{P}$.

As a result, the relation between $\mathbf{S}$ and $\mathbf{X}$ can be written in matrix form as 
\begin{equation} 
{\bf{S}} = {{\bf{A}}^{\rm H}}{\bf{XF}}_M^{\rm H} .
		\label{eq: relation between X and X in matrix form}
\end{equation}
Similarly to AFDM, the proposed ADDM needs to add a chirp-periodic prefix (CPP) to combat multipath propagation and make the channel seemingly lie in a periodic domain\cite{bemani2023affine}. However, due to the different signal periodicity of DAFT in (\ref{eq: periodicity of Sm}), a chirp-periodic prefix (CPP) is used here. The CPP can be written as
\begin{equation} 
{{\bf{S}}_{\mathrm{cp}}}\left[ {n,k} \right] = {\bf{S}}\left[ {N + n,k} \right]{e^{ - j2\pi {c_1}\left( {{N^2} + 2Nn } \right)}} ,
		\label{eq: explanation of CPP}
\end{equation}
where $n = - {N_{\mathrm{cp}}}, \ldots , - 1$.
 ${{N}_{\mathrm{cp}}}$ is the length of CPP.

Let $\mathbf{\tilde S}$ denote the matrix after adding CPP, and then $\mathbf{\tilde S}$ can be obtained as
\begin{equation} \label{eq: relation between the tdomain data added CP x' and x in matrix form}
{\mathbf{\tilde S}} = {\bf{T}}_{\mathrm{cp}}^{{N_{\mathrm{cp}}},{c_1}}{\bf{S}} ,
\end{equation}
where $\mathbf{T}_{\mathrm{cp}}^{{N_{\mathrm{cp}}},{{c}_{1}}}\in {{\mathbb{C}}^{\left( N+{N_{\mathrm{cp}}} \right)\times N}}$, whose elements are determined by parameters ${N_{\mathrm{cp}}}$ and ${{c}_{1}}$, i.e., 
\begin{equation} 
{\bf{T}}_{\mathrm{cp}}^{{N_{\mathrm{cp}}},{c_1}} = {\left[ {\begin{array}{*{20}{c}}
			{{{\left[ {\begin{array}{*{20}{c}}
								{{{\bf{O}}_{{N_{\mathrm{cp}}} \times (N - {N_{\mathrm{cp}}})}}|{\bf{\Psi }}}
						\end{array}} \right]}^{\rm T}}|{{\bf{I}}_N}}
	\end{array}} \right]^{\rm T}} .
	\label{eq: explanation of adding CP matrix}
\end{equation}
$\left[ {{\bf{a}}|{\bf{b}}} \right]$ denotes arranging the matrices ${\bf{a}}$ and ${\bf{b}}$ horizontally.
${\bf{\Psi }}$ is a diagonal matrix which can be explained as
\begin{equation} 
\mathbf{\Psi }=\mathrm{diag}({{e}^{-j2\pi {{c}_{1}}\left( {{N}^{2}}+2Nn-2N{{N}_{\mathrm{cp}}} \right)}},n=0,1,\ldots ,{{N}_{\mathrm{cp}}}-1) .
		\label{eq: explanation of PHI in adding CP matrix}
\end{equation}
Consequently, the relation between $\mathbf{\tilde S}$ and  $\mathbf{X}$ is
\begin{equation} 
\mathbf{{\tilde S}}=\mathbf{T}_{\mathrm{cp}}^{{{N}_{\mathrm{cp}}},{{c}_{1}}}{{\mathbf{A}}^{\rm H}}\mathbf{XF}_{M}^{\rm H}.
		\label{eq: relation between x' and X in matrix form}
\end{equation}
After parallel to serial conversion (P/S), the transmitted signal in the time domain can be written as
\begin{equation} \label{eq:Tx_1}
{\mathrm{vec}}\left( \mathbf{{\tilde S}} \right) = {\mathrm{vec}}\left( {{\bf{T}}_{\mathrm{cp}}^{{{N}_{\mathrm{cp}}},{c_1}}{{\bf{A}}^{\rm H}}{\bf{XF}}_M^{\rm H}} \right) ,
\end{equation}
where ${\mathrm{vec}}\left( \cdot \right)$ denotes stacking columns of a matrix into a column vector. 

\vspace{0.5ex}
\noindent\emph{(2) Demodulation}
\vspace{0.5ex}

After transmission over the channel, serial to parallel conversion (S/P) and discarding CPP at the receiver, the received signal matrix in the T-D domain is given by 
\begin{equation} 
{\bf{R}} = \sum\limits_{i = 1}^P {{{\tilde h}_i}{{\bf{\Gamma }}_{\mathrm{cpp}_i}}{{\bf{\Delta }}_{{f_{1i}}}}{{\bf{\Pi }}^{{l_i}}}{\bf{S}}{{\bf{\Delta }}_{{f_{2i}}}}}  + {\bf{W}} ,
	\label{eq: relation between R and x in matrix form}
\end{equation}
where ${{\tilde{h}}_{i}}={{h}_{i}}{{e}^{-j2\pi {{f}_{\mathrm{d},i}}{{\tau }_{i}}}}$. ${f}_{\mathrm{d},i}$ and ${\tau }_{i}$ are the Doppler shift and the time delay of the $i$-th path, respectively. ${{\mathbf{\Gamma }}_{\mathrm{cpp}_i}}\in {{\mathbb{C}}^{N\times N}}$ is a diagonal matrix which can be explained as
\begin{equation} 
	{{\mathbf{\Gamma }}_{\mathrm{cpp}_i}}=\mathrm{diag}(\left\{ \begin{array}{*{35}{l}}
		{{e}^{-j2\pi {{c}_{1}}\left( {{N}^{2}}-2N\left( {{l}_{i}}-n \right) \right)}} & n<{{l}_{i}}  \\
		1 & n\ge {{l}_{i}}  \\
	\end{array} \right.),
	\label{eq: explanation of CPP affect of the channel in matrix form}
\end{equation}
with $n=0,1,\ldots ,N-1$. The structure of it
depends on the mathematical form of CPP shown in Eq. (\ref{eq: explanation of CPP}).
$\mathbf{\Pi }$ is the forward cyclic-shift matrix
\begin{equation} 
	\mathbf{\Pi }={{\left[ \begin{matrix}
				0 & \ldots  & 0 & 1  \\
				1 & \ldots  & 0 & 0  \\
				\vdots  & \ddots  & \ddots  & \vdots   \\
				0 & \ldots  & 1 & 0  \\
			\end{matrix} \right]}_{N\times N}} ,
	\label{eq: explanation of delay affect of the channel in matrix form}
\end{equation}
which represents the cyclic shift of ${\bf{s}}$ on the column. ${{\bf{\Delta }} _{{f_{1i}}}} \buildrel \Delta \over = \mathrm{diag}\left( {{e^{j2\pi {f_i}n}}} , n = 0, \ldots ,N - 1\right)$ and ${{\bf{\Delta }} _{{f_{2i}}}} \buildrel \Delta \over =  \mathrm{diag}\left( {{e^{j2\pi {f_i}m{N_\mathrm{s}}}}}, m = 0, \ldots ,M - 1 \right)$ with ${{N}_{\mathrm{s}}}=N+{{N}_{\mathrm{cp}}}$. They denote the Doppler shift on the column and the row, respectively. $\mathbf{W}\in {{\mathbb{Z}}^{N\times M}}$ is an additive Gaussian noise matrix with power spectral density $\sigma _n^2$. 

Then, DAFT is performed for each column of $\mathbf{R}$, and the symbol matrix in the T-A domain can be obtained as 
\begin{equation} 	\label{eq: relation between Y and R in elements}
	\mathbf{Y}\left[ m',k \right]=\frac{1}{\sqrt{N}}\sum\limits_{n=0}^{N-1}{\mathbf{R}\left[ n,k \right]}{{e}^{-j2\pi \left( {{c}_{1}}{{n}^{2}}+{{c}_{2}}m{{'}^{2}}+\frac{m'n}{N} \right)}} .			
\end{equation}
After that, DFT is used for each row of $\mathbf{Y}$, and the received symbol matrix in the A-D domain is given by 
\begin{equation} 
{\bf{Z}}\left[ {m',q} \right] = \frac{1}{{\sqrt M }}\sum\limits_{k = 0}^{M - 1} {\bf{Y}} \left[ {m',k} \right]{e^{ - j2\pi \frac{{kq}}{M}}} .				
\end{equation}
In matrix form, the received symbol matrix in the A-D domain can be written as
\begin{equation} \label{eq:Z_1}
{\bf{Z}} =  \sum\limits_{i = 1}^P {{\bf{A}}{{\widetilde h}_i{{\bf{\Gamma }}_{\mathrm{cpp}_i}}{{\bf{\Delta }} _{{f_{1i}}}}{{\bf{\Pi }}^{{l_i}}}{\bf{S}}} {{\bf{\Delta }} _{{f_{2i}}}}{{\bf{F}}_M}} + {\bf{A}\bf{W}}{{\bf{F}}_M} .		
\end{equation}

\vspace{0.5ex}
\noindent\emph{(3) Input-output relation in the A-D domain}
\vspace{0.5ex}

Based on Eq. (\ref{eq:Z_1}), the input-output relation in the A-D domain can be expressed as
\begin{align} 
	{\bf{Z}} &= \sum\limits_{i = 1}^P {{\bf{A}}{{\widetilde h}_i{{\bf{\Gamma }}_{\mathrm{cpp}_i}}{{\bf{\Delta }} _{{f_{1i}}}}{{\bf{\Pi }}^{{l_i}}}{{\bf{A}}^{\rm H}}{\bf{XF}}_M^{\rm H}} {{\bf{\Delta }} _{{f_{2i}}}}{{\bf{F}}_M}} + {\bf{\tilde W}} \nonumber \\	
	& = \sum\limits_{i = 1}^P {{{\widetilde h}_i}{{\bf{H}}_{\mathrm{A},i}}{\bf{X}}} {{\bf{H}}_{\mathrm{D},i}} + {\bf{\tilde W}} ,
	\label{eq: relation between Z and X in matrix form}
\end{align}
where ${{\bf{H}}_{\mathrm{A},i}} \buildrel \Delta \over = {\bf{A}}{{\bf{\Gamma }}_{\mathrm{cpp}_i}}{{\bf{\Delta }} _{{f_{1i}}}}{{\bf{\Pi }}^{{l_i}}}{{\bf{A}}^\mathrm{H}}$, ${{\bf{H}}_{\mathrm{D},i}} \buildrel \Delta \over = {\bf{F}}_M^\mathrm{H}{{\bf{\Delta }} _{{f_{2i}}}}{{\bf{F}}_M}$, and ${\bf{\tilde W}} = {\bf{AW}}{{\bf{F}}_M}$. 
After parallel to serial conversion (P/S), the input-output relation in the A-D domain is shown as
\begin{align} \label{eq:Z_2}
	{\rm{vec}}\left( {\bf{Z}} \right) &= \sum\limits_{i = 1}^P {{{\widetilde h}_i} {\rm{vec}}\left({{\bf{H}}_{\mathrm{A},i}}{\bf{X}}  {{\bf{H}}_{\mathrm{D},i}} \right) }+ {\rm{vec}}\left({\bf{\tilde W}}\right) \nonumber \\	
	& = \sum\limits_{i = 1}^P {{{\widetilde h}_i} {{\bf{H}}_{{\rm eff},i}} {\rm{vec}}\left({\bf{X}}   \right) }+ {\rm{vec}}\left({\bf{\tilde W}}\right) ,	
\end{align}
where $\mathbf{H}_{\mathrm{eff},i}=\mathbf{H}_{\mathrm{D},i}^T\otimes\mathbf{H}_{\mathrm{A},i}^T$, which denotes the effective channel in the affine-Doppler domain. 

Through mathematical derivation, ${{\mathbf{H}}_{\mathrm{D},i}}\left[ p,q \right]$ can be obtained as
\begin{equation} \label{eq: Hi(p,q)}
	{{\bf{H}}_{\mathrm{D},i}}\left[ {p,q} \right] = \frac{1}{M}{{\bf{Q}}_i}\left[ {p,q} \right] ,
\end{equation}
where 
\begin{align} \label{eq:Q}
	{{{\bf{Q}}_i}\left[ {p,q} \right]}&{ \buildrel \Delta \over = \sum\limits_{k = 0}^{M - 1} {{e^{j\frac{{2\pi }}{M}\left( {M{N_\mathrm{s}}{f_i} + p - q} \right)k}}} } \nonumber \\
	&{=} { \frac{{{e^{j2\pi \left( {M{N_\mathrm{s}}{f_i} + q - p} \right)}} - 1}}{{{e^{j\frac{{2\pi }}{M}\left( {M{N_\mathrm{s}}{f_i} + q - p} \right)}} - 1}}} .	
\end{align}
According to \cite{bemani2023affine}, ${{\mathbf{H}}_{\mathrm{A},i}}\left[ m',m \right]$ is given by
\begin{equation} \label{eq: explanation of Hi(m',m)}
	{{\mathbf{H}}_{\mathrm{A},i}}\left[ m',m \right]=\frac{1}{N}{{\bf{E}}_{i}}\left[ m',m \right]{{\bf{K}}_{i}}\left[ m',m \right],	
\end{equation}
where
\begin{equation} \label{eq:E}
	{{\bf{E}}_{i}}\left[m',m \right]\triangleq {{e}^{j\frac{2\pi }{N}\left[ N{{c}_{1}}l_{i}^{2}-m{{l}_{i}}+N{{c}_{2}}\left( {{m}^{2}}-m{{'}^{2}} \right) \right]}},
\end{equation}
and 
\begin{align} \label{eq:K}
	{{\bf{K}}_{i}}\left[ m',m \right]\; &{=} \sum\limits_{n=0}^{N-1}{{{e}^{j\frac{2\pi }{N}\left( N{{f}_{i}}-2N{{c}_{1}}{{l}_{i}}+m-m' \right)n}}} \nonumber \\
	&{=} \frac{{{e}^{j2\pi \left( N{{f}_{i}}-2N{{c}_{1}}{{l}_{i}}+m-m' \right)}}-1}{{{e}^{j\frac{2\pi }{N}\left( N{{f}_{i}}-2N{{c}_{1}}{{l}_{i}}+m-m' \right)}}-1} .						
\end{align}

Let ${{\nu}_{i}}\triangleq N{{f}_{i}}={{\alpha }_{i}}+{{a}_{i}}$, which denotes the Doppler shift normalized to the subcarrier spacing $\Delta _\mathrm{f}$, and ${{\nu}_i'} \buildrel \Delta \over = {N_\mathrm{s}}{f_i} = {\beta _i} + {b_i}$. ${\alpha _i} {\in} \left[ { - {\alpha _{\max }},{\alpha _{\max }}} \right]$ and ${{\beta }_{i}}$ are the integral part of ${{v}_{i}}$ and ${v_i}'$, respectively, whereas ${{a}_{i}}$ and ${{b}_{i}}$ are the fractional part satisfying $-\frac{1}{2}<{{a}_{i}}\le \frac{1}{2}$ and $-\frac{1}{2}<{{b}_{i}}\le \frac{1}{2}$. From Eq. (\ref{eq:Q}) and Eq. (\ref{eq:K}), the values of ${{\bf{Q}}_{i}}\left[ p,q \right]$ and ${{\bf{K}}_{i}}\left[ m',m \right]$ depend on ${{\nu}_{i}'}$ and ${{\nu}_i}$, respectively. It is assumed that ${{c}_{1}}$ is chosen such that $2N{{c}_{1}}{{l}_{i}}$ is an integer.

${{{\bf{Q}}_i}\left[ {p,q} \right]}$ is a function of $b_i$.
Let $\nu _i '' = Mb_i$. 
\romannumeral 1) Integer ${{\nu}_{i}''}$: When ${{\nu}_{i}''}$ is integer, i.e., $\left\lfloor Mb_i \right\rceil = Mb_i$, $\left\lfloor \cdot \right\rceil$ denotes the round function, Eq. (\ref{eq:Q}) is equal to
\begin{equation} \label{eq: Qi(p,q) if bi=0}
	{{{\bf{Q}}_i}\left[ {p,q} \right]} = \left\{ {\begin{array}{*{20}{l}}
			M,&{p = {{\left\langle {q + \left\lfloor Mb_i \right\rceil} \right\rangle }_M}},\\
			0,&{otherwise},
	\end{array}} \right. 	
\end{equation}
where ${{\left\langle \cdot  \right\rangle }_{M}}$ denotes the modulo $M$ operation.
\romannumeral 2) Fractional ${{\nu}_{i}''}$: When ${{\nu}_{i}''}$ is fractional, $\left| {{{\bf{Q}}_i}\left[ {p,q} \right]} \right|$ can be obtained as
\begin{align} 	\label{eq:abs_K}
	\left| {{{\bf{Q}}_i}\left[ {p,q} \right]} \right| {=} \left| \frac{{{e}^{jM\phi }}\left( {{e}^{jM\phi }}-{{e}^{-jM\phi }} \right)}{{{e}^{j\phi }}\left( {{e}^{j\phi }}-{{e}^{-j\phi }} \right)} \right| {=}  \frac{\left| \sin M\phi  \right|}{\left| \sin \phi  \right|} ,
\end{align}
where $\phi =  \frac{{\pi }}{M}\left( {Mb_i + q - p} \right)$. It means that the magnitude of ${{{\bf{Q}}_i}\left[ {p,q} \right]}$ has a peak at
$p = {{\left\langle {q + \left\lfloor Mb_i \right\rceil} \right\rangle }_M}$ and decreases as $q$ moves away from the peak. 
We consider that ${{{\bf{Q}}_i}\left[ {p,q} \right]}$ is non-zero for $2{{k}_{\mathrm{f}}}+1$ values for the $q$-column corresponding to an interval centered at ${\left\langle {q + \left\lfloor Mb_i \right\rceil} \right\rangle _M}$. ${k}_\mathrm{f}$ is a parameter which is chosen in such a way that when $\left| p - {{\left\langle {q + \left\lfloor Mb_i \right\rceil} \right\rangle }_M} \right| > {k_\mathrm{f}}$, the value of $\left| {{{\bf{Q}}_i}\left[ {p,q} \right]} \right|$ is smaller than a threshold. 

${{\bf{K}}_{i}}\left[ {m',m} \right]$ is a function of $\alpha _i$.
\romannumeral 1) Integer ${{\nu}_{i}}$: When ${{\nu}_{i}}$ is integer, i.e., ${{a}_{i}}=0$ and ${{\nu}_{i}} = \alpha _i$, Eq. (\ref{eq:K}) can be rewritten as
\begin{equation} \label{eq:K_ai_0}
{{\bf{K}}_{i}}\left[ {m',m} \right] = \left\{ {\begin{array}{*{20}{l}}
		N,&{m = {{\left\langle {m' + 2N{c_1}{l_i} - {{\alpha}_i}} \right\rangle }_N}},\\
		0,&{otherwise}.
\end{array}} \right.						
\end{equation}
\romannumeral 2) Fractional ${{\nu}_{i}}$: When ${{\nu}_{i}}$ is fractional, according to \cite{bemani2023affine}, $\left| {{\bf{K}}_{i}}\left( m',m \right) \right|$ can be obtained as
\begin{align} 
	\left| {{\bf{K}}_{i}}\left[ m',m \right] \right|\; &{=} \left| \frac{{{e}^{jN\theta }}\left( {{e}^{jN\theta }}-{{e}^{-jN\theta }} \right)}{{{e}^{j\theta }}\left( {{e}^{j\theta }}-{{e}^{-j\theta }} \right)} \right| {=}  \frac{\left| \sin N\theta  \right|}{\left| \sin \theta  \right|} ,
	\label{eq:|Ki(m',m)|}
\end{align}
where $\theta =  \frac{\pi }{N}\left( {{{\nu}_i} - 2N{c_1}{l_i} + m - m'} \right)$. It means that the magnitude of ${{\bf{K}}_{i}}\left[ m',m \right]$ has a peak at
$m = {{\left\langle {m' + 2N{c_1}{l_i} - {{\alpha}_i}} \right\rangle }_N}$ and decreases as $m$ moves away from ${\left\langle m' + 2N{c_1}{l_i} - {{\alpha}_i} \right\rangle }_N$\cite{bemani2023affine}. We consider that $\left| {{\bf{K}}_{i}}\left[ m',m \right] \right|$ is non-zero only for $2{{k}_{\mathrm{a}}}+1$ values of $m$ corresponding to an interval centered at ${\left\langle {m' + 2N{c_1}{l_i} - {{\alpha}_i}} \right\rangle _N}$\cite{bemani2023affine}. ${k}_\mathrm{a}$ is a parameter which is chosen in such a way that when $\left| m - {\left\langle {m' + 2N{c_1}{l_i} - {{\alpha}_i}} \right\rangle _N} \right| > {k_\mathrm{a}}$, the value of $\left| {{\bf{K}}_{i}}\left[ m',m \right] \right|$ is smaller than a threshold.

\begin{figure}[htbp]
	\centering
	\begin{minipage}{0.7\linewidth}
		\centering
		\includegraphics[width=0.9\linewidth]{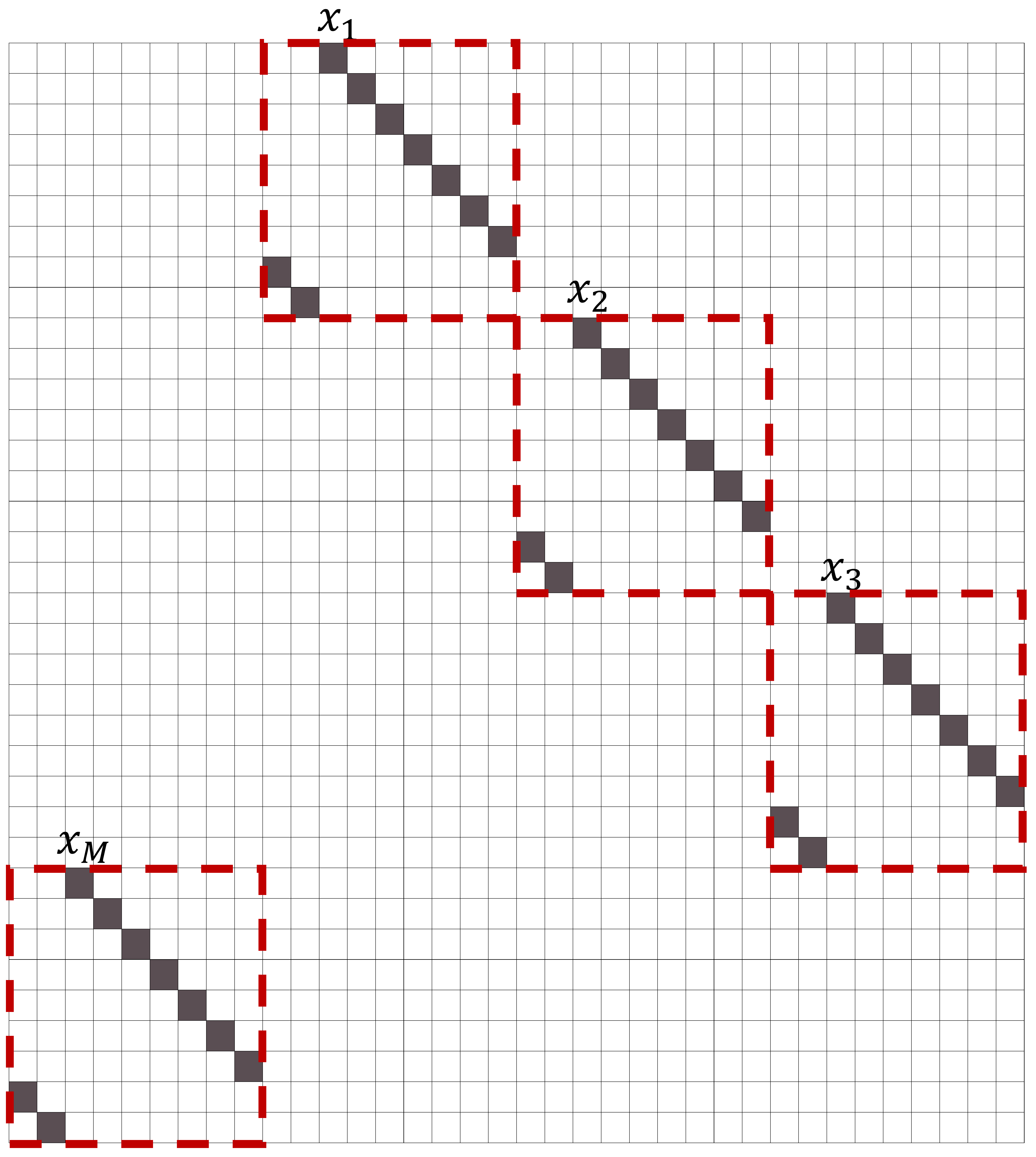}
		\caption{Structure of $\mathbf{H}_{\mathrm{eff},i}$.}
		
		\label{fg:channel matrix integer}
	\end{minipage}
\end{figure}

In conclusion, when ${{\nu}_{i}}$ and ${{\nu}_i}''$ are both integers, the input-output relation in the A-D domain is shown as
\begin{align} \label{eq:In_out_IntCase}
	{\bf{Z}}\left[ {m',q} \right] &= \sum\limits_{i = 1}^P {{{\tilde h}_i}{e^{j\frac{{2\pi }}{N}\left[ {N{c_1}l_i^2 - m{l_i}  + N{c_2}\left( {{m^2} - m{'^2}} \right)} \right]}}{\bf{X}}\left[ {m,p} \right]} \nonumber \\
	 & \qquad \qquad + {\bf{\tilde W}}\left[ {m',q} \right] ,	
\end{align}
where ${m = {{\left\langle {m' + 2N{c_1}{l_i} - {{\alpha}_i}} \right\rangle }_N}}$ and ${p = {{\left\langle {q + \left\lfloor Mb_i \right\rceil} \right\rangle }_{{M}}}}$. 
When ${{\nu}_{i}}$ or ${{\nu}_i}''$ is fractional, the input-output relation in the A-D domain is given by Eq. (\ref{eq:In_out_fraCase}) at the top of next page. For a given channel of a single path with $M=4$, $N=9$, $2Nc_1=12$, $l_i=1$, $b_i=\frac14$, ${{\nu}_{i}}=\alpha_i=1$ and ${{\nu}_i}''=Mb_i=1$,
the structure of $\mathbf{H}_{\mathrm{eff},i}=\mathbf{H}_{\mathrm{D},i}^T\otimes\mathbf{H}_{\mathrm{A},i}^T$ is shown as Fig. \ref{fg:channel matrix integer}  where $x_1=\left\langle Mb_i\right\rangle_MN+\left\langle2Nc_1l_i-\alpha_i\right\rangle_N$, and $\left\langle x_{m}-x_{m-1}\right\rangle_{MN}=N$, which reveals that the effective channel of ADDM in A-D domain is still sparse.


\small
\begin{figure*}[ht]
	\begin{align}\label{eq:In_out_fraCase}		
		{\bf{Z}}\left[ {m',q} \right] &= \frac{1}{{NM}}\sum\limits_{i = 1}^P {{{\tilde h}_i}}  \sum\limits_{m = {{\left\langle {m' + 2N{c_1}{l_i} - {{\alpha}_i}}- {k_\mathrm{a}} \right\rangle }_N} }^{{{\left\langle {m' + 2N{c_1}{l_i} - {{\alpha}_i}} + {k_\mathrm{a}} \right\rangle }_N} } \ \ {\sum\limits_{p = {{\left\langle {q + \left\lfloor Mb_i \right\rceil - {k_\mathrm{f}}} \right\rangle }_M}}^{ {{\left\langle {q + \left\lfloor Mb_i \right\rceil} + {k_\mathrm{f}} \right\rangle }_M}} {{e^{j\frac{{2\pi }}{N}\left[ {N{c_1}l_i^2 - m{l_i} + N{c_2}\left( {{m^2} - m{'^2}} \right)} \right]}}\frac{{{e^{j2\pi \left( {{{\nu}_i} - 2N{c_1}{l_i} + m - m'} \right)}} - 1}}{{{e^{j\frac{{2\pi }}{N}\left( {{{\nu}_i} - 2N{c_1}{l_i} + m - m'} \right)}} - 1}}} } \nonumber \\
		&\qquad \qquad \qquad \qquad \qquad \qquad \qquad \qquad \qquad \qquad \qquad \qquad \times \frac{{{e^{j2\pi \left( {{{\nu}_i''} + q - p} \right)}} - 1}}{{{e^{j\frac{{2\pi }}{M}\left( {{{\nu}_i''} + q - p} \right)}} - 1}}{\bf{X}}\left[ {m,p} \right] + {\bf{\tilde W}}\left[ {m',q} \right].
	\end{align}
\end{figure*}
\normalsize
\section{Compatibility Analyses of ADDM}


In this subsection, we show that 
the proposed ADDM can subsume the existing typical multicarrier waveforms or chirp-based waveforms as a particular case.
Firstly, a general form of transmitted ADDM waveform is given by adding a reduced cyclic prefix (RCP) matrix ${{\bf{T}}_{\mathrm{rcp}}}$ based on Eq. (\ref{eq:Tx_1}), which is shown as
\begin{equation} \label{eq:general_form_ADDM}
	{\bf{\tilde x}}_{\mathrm{g}} = {\bf{T}}_{\mathrm{rcp}}^{{N_{\mathrm{rcp}}}}{\rm{vec}}\left( {{\bf{T}}_{{\rm{cp}}}^{{{N}_{{\rm{cp}}}},{c_1}}{{\bf{T}}_{{{\rm{F}}}}}{\bf{T}}_{{{\rm{V}}}}^{{N_\mathrm{v}}}{\bf{X}}{{\bf{T}}_{\rm{B}}}} \right) ,
\end{equation}
where ${{N}_{\mathrm{rcp}}}$ is the length of reduced cyclic prefix (RCP), and
\begin{equation} \label{eq: explanation of adding RCP matrix TRcp}
	{\bf{T}}_{\mathrm{Rcp}}^{{N_{\mathrm{rcp}}}} = {\left[ {{{\left[ {{\bf{O}}|{{\bf{I}}_{{N_{\mathrm{rcp}}}}}} \right]}^T}|{{\bf{I}}_{(N + {N_{\mathrm{cp}}})M}}} \right]^\mathrm{T}} ,
\end{equation}
where $\mathbf{O}$ is the null matrix with ${{N}_{\mathrm{rcp}}}$ rows and $\left( N+{N_{\mathrm{cp}}} \right)M-{{N}_{\mathrm{rcp}}}$ columns. ${{\mathbf{T}}_{{{\mathrm{F}}}}}\in {{\mathbb{C}}^{N\times N}}$ is the forward matrix, ${{\mathbf{T}}_{\mathrm{B}}}\in {{\mathbb{C}}^{M\times M}}$ denotes the backward matrix, and
$\mathbf{T}_{{{\mathrm{V}}}}^{{{N}_{\mathrm{v}}}}\in {{\mathbb{C}}^{N\times {{N}_{\mathrm{v}}}}}$ is used to adjust the number of valid symbols arranged in the affine domain, which is given by
\begin{equation} \label{eq: explanation of TF1Nr}
	{\bf{T}}_{{\mathrm{V}}}^{{N_\mathrm{v}}} = {\left[ {\begin{array}{*{20}{c}}
				{{{\bf{I}}_{{N_\mathrm{v}}}}|{{\bf{O}}_{{N_\mathrm{v}} \times (N - {N_\mathrm{v}})}}}
		\end{array}} \right]^\mathrm{T}},
\end{equation}
where $N_\mathrm{v}$ is the number of valid symbols arranged in the affine domain.

According to the general form of ADDM waveform shown in Eq. (\ref{eq:general_form_ADDM}), setting different matrices (i.e., ${\bf{T}}_{\mathrm{rcp}}^{{N_{\mathrm{rcp}}}}$, ${\bf{T}}_{{\rm{cp}}}^{{{L}_{{\rm{cp}}}},{c_1}}$, ${{\bf{T}}_{{{\rm{F}}}}}$, ${\bf{T}}_{{{\rm{V}}}}^{{N_\mathrm{v}}}$, and ${{\bf{T}}_{\rm{B}}}$) can obtain not only the new ADDM waveform, but also the existing CP-AFDM, CP-OFDM, CP-OCDM, CP-FDDM, CP-OTFS, RCP-OTFS,  and linear frequency modulation (LFM) waveforms.
The relationship between waveform and corresponding matrixes is listed in Table \ref{tb:wave_matrix}. This means that AFDM, OTFS, OCDM, OFDM, FDDM, and LFM waveforms can be regarded as special cases of ADDM. Thus, ADDM can potentially have the advantages of these waveforms simultaneously.

\begin{table}[htbp] 
	\centering 
	\caption{The relation between waveforms and corresponding matrices.}
	\resizebox{3.5in}{!}{ 
		\setstretch{1.5} 
		\begin{tabular}{|c|c|c|c|c|c|} \hline 
			\diagbox{Waveform}{Matrix} & $\mathbf{T}_{\mathrm{rcp}}^{{{N}_{\mathrm{rcp}}}}$ & $\mathbf{T}_{\mathrm{cp}}^{{{N}_{\mathrm{cp}}},{{c}_{1}}}$ & ${{\mathbf{T}}_{{{\mathrm{F}}}}}$ & $\mathbf{T}_{{{\mathrm{V}}}}^{{{N}_{\mathrm{v}}}}$ & ${{\mathbf{T}}_{\mathrm{B}}}$  \\ \hline
			CP-ADDM & $\mathbf{T}_{\mathrm{rcp}}^{{0}}$ & $\mathbf{T}_{\mathrm{cp}}^{{{N}_{\mathrm{cp}}},{{c}_{1}}}$ & ${{\mathbf{A}}_{c_1,c_2}^{\mathrm{H}}}$ & $\mathbf{T}_{{{\mathrm{V}}}}^{N}$ & $\mathbf{F}_{M}^{\mathrm{H}}$ \\ \hline
			CP-AFDM &	$\mathbf{T}_{\mathrm{rcp}}^{{0}}$ & $\mathbf{T}_{\mathrm{cp}}^{{{N}_{\mathrm{cp}}},{{c}_{1}}}$ & ${{\mathbf{A}}_{c_1,c_2}^{\mathrm{H}}}$ & $\mathbf{T}_{{{\mathrm{V}}}}^{N}$ & $\mathbf{F}_{1}^{\mathrm{H}}$ \\ \hline
			CP-OFDM & $\mathbf{T}_{\mathrm{rcp}}^{{0}}$ & $\mathbf{T}_{\mathrm{cp}}^{{{N}_{\mathrm{cp}}},0}$ & ${{\mathbf{A}}_{0,0}^{\mathrm{H}}}$ & $\mathbf{T}_{{{\mathrm{V}}}}^{N}$ & $\mathbf{F}_{1}^{\mathrm{H}}$ \\ \hline  
			CP-OCDM & $\mathbf{T}_{\mathrm{rcp}}^{{0}}$ & $\mathbf{T}_{\mathrm{cp}}^{{{N}_{\mathrm{cp}}},{1 \mathord{\left/
						{\vphantom {1 2N}} \right.
						\kern-\nulldelimiterspace} 2N}}$ & ${{\mathbf{A}}_{{1 \mathord{\left/
							{\vphantom {1 2N}} \right.
							\kern-\nulldelimiterspace} 2N},{1 \mathord{\left/
							{\vphantom {1 2N}} \right.
							\kern-\nulldelimiterspace} 2N}}^{\mathrm{H}}}$ & $\mathbf{T}_{{{\mathrm{V}}}}^{N}$ & $\mathbf{F}_{1}^{\mathrm{H}}$ \\ \hline
			CP-FDDM & $\mathbf{T}_{\mathrm{rcp}}^{{0}}$ & $\mathbf{T}_{\mathrm{cp}}^{{{N}_{\mathrm{cp}}},{{c}_{1}}}$ & ${{\mathbf{A}}_{0,0}^{\mathrm{H}}}$ & $\mathbf{T}_{{{\mathrm{V}}}}^{N}$ & $\mathbf{F}_{M}^{\mathrm{H}}$ \\ \hline			
			CP-OTFS & $\mathbf{T}_{\mathrm{rcp}}^{{0}}$ & $\mathbf{T}_{\mathrm{cp}}^{{{N}_{\mathrm{cp}}},0}$ & ${{\mathbf{I}}_{N}}$ & $\mathbf{T}_{{{\mathrm{V}}}}^{N}$ & $\mathbf{F}_{M}^{\mathrm{H}}$  \\ \hline
			RCP-OTFS & $\mathbf{T}_{\mathrm{rcp}}^{{{N}_{\mathrm{rcp}}}}$ & $\mathbf{T}_{\mathrm{cp}}^{{0},0}$ & ${{\mathbf{I}}_{N}}$ &	$\mathbf{T}_{{{\mathrm{V}}}}^{N}$ & $\mathbf{F}_{M}^{\mathrm{H}}$  \\ \hline
			LFM & $\mathbf{T}_{\mathrm{rcp}}^{{0}}$ & $\mathbf{T}_{\mathrm{cp}}^{{0},0}$ & ${{\mathbf{A}}_{c_1,0}^{\mathrm{H}}}$ & $\mathbf{T}_{{{\mathrm{V}}}}^{1}$ & $\mathbf{F}_{1}^{\mathrm{H}}$  \\ \hline	
		\end{tabular}
	}\label{tb:wave_matrix}
\end{table}

\section{Simulation Results}

In this section, numerical results based on Monte Carlo simulations are presented. We compare the BER performances of our proposed ADDM waveform and the existing AFDM and OTFS waveforms versus signal-to-noise ratio (SNR) in doubly selective channels \cite{bemani2023affine}. In our simulation, the carrier frequency $f_\mathrm{c} = 24$ GHz, and the bandwidth $B=7.68$ MHz. The QPSK symbols are transmitted. We let the symbols of ADDM/AFDM/OTFS have the identical bandwidth $B=7.68$ MHz and time interval $T=0.2667$ ms to enjoy the same delay and Doppler resolutions, i.e., $N = 128$ and $M = 16$ for ADDM and OTFS, and $N = 2048$ and $M = 1$ for AFDM. The number of CP (CPP) $N_{\mathrm{cp}}=4$ for all waveforms. The value of $c_1$ is set to be 0.1211 for ADDM and AFDM. Ideal channel information is assumed to be obtained, and MMSE is performed for all waveforms\cite{bemani2023affine}.

We consider a channel with $P=3$ paths, whose maximum integer part of the normalized Doppler shift is $\alpha _{\mathrm{max}}$ = 2\cite{bemani2023affine}. The complex gain of the $i$-th path ${{h}_{i}}$ is set to be independent complex Gaussian random variables with zero mean and $1/P$ variance. We use ``Case I'' to represent the case that the three paths have different delays $l=\left[0,1,2\right]$, and use ``Case II'' to denote the case that the three paths have identical delay $l=\left[1,1,1\right]$\cite{Raviteja2018Interference,yin2022design}. In both cases, each path has a different Doppler shift generated by the Jakes model, i.e., $\nu _i = \alpha _{\mathrm{max}} \mathrm{cos}(\theta _i)$, where $\theta _i$ is uniformly distributed over $\left[-\pi,\pi\right]$\cite{bemani2023affine}.

\begin{figure}[htbp]
	\centering
	\includegraphics[width=3in]{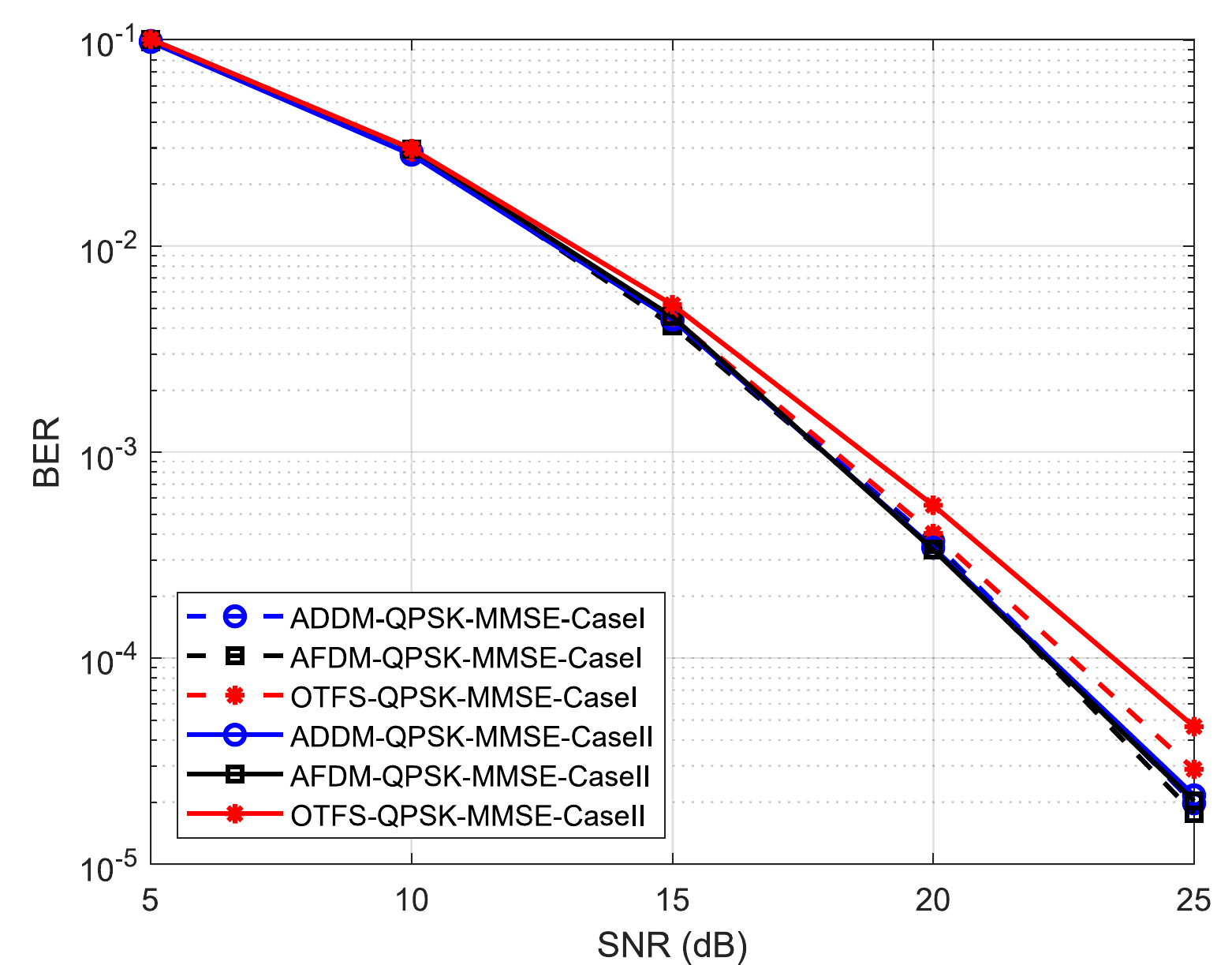}
	\caption{The BER performances of ADDM, AFDM and OTFS.
		\label{fg:Normalized_SNR_to_number_of_transmitter}}
\end{figure}

We can see from Fig. \ref{fg:Normalized_SNR_to_number_of_transmitter} that for ``Case I'', three waveforms achieve comparable BER performance, and for ``Case II'', our proposed ADDM and AFDM have almost the same BER performance, outperforming OTFS. The reason is that the unambiguous Doppler of OTFS is limited by the subcarrier spacing $\Delta _\mathrm{f}$. When three paths have an identical delay, and the maximum Doppler of the path is $\alpha _{\mathrm{max}}\Delta _\mathrm{f}$ = $2\Delta _\mathrm{f}$, the effective channel matrices of three paths may overlap in the D-D domain, which causes OTFS not to obtain the full diversity. However, for our proposed ADDM and AFDM, the unambiguous Doppler is several times the subcarrier spacing, which results in our ADDM and AFDM achieving full diversity and thus outperforming OTFS. 

\section{Conclusion}

This paper introduced a new orthogonal multicarrier waveform, namely ADDM, which modulates information symbols in the A-D domain based on a two-dimensional (2D) DAFT and DFT. ADDM provided a generic framework and subsumed existing OFDM, FDDM, OTFS and AFDM as a particular case. The input-output relation of ADDM in the A-D domain was derived, revealing that ADDM achieves full diversity order of doubly selective channels. Numerical results showed that our proposed ADDM and AFDM have almost the same BER performance, outperforming OTFS in high-mobility scenarios. 

\section*{Acknowledgment}

This work was supported by the National Natural Science Foundation of China under Grant 61971025, 62331002 and 62071026.

\bibliographystyle{IEEEtran}
\bibliography{IEEEabrv,IEEE_RadarConf}

\begin{thebibliography}{10}
\providecommand{\url}[1]{#1}
\csname url@samestyle\endcsname
\providecommand{\newblock}{\relax}
\providecommand{\bibinfo}[2]{#2}
\providecommand{\BIBentrySTDinterwordspacing}{\spaceskip=0pt\relax}
\providecommand{\BIBentryALTinterwordstretchfactor}{4}
\providecommand{\BIBentryALTinterwordspacing}{\spaceskip=\fontdimen2\font plus
\BIBentryALTinterwordstretchfactor\fontdimen3\font minus
  \fontdimen4\font\relax}
\providecommand{\BIBforeignlanguage}[2]{{%
\expandafter\ifx\csname l@#1\endcsname\relax
\typeout{** WARNING: IEEEtran.bst: No hyphenation pattern has been}%
\typeout{** loaded for the language `#1'. Using the pattern for}%
\typeout{** the default language instead.}%
\else
\language=\csname l@#1\endcsname
\fi
#2}}
\providecommand{\BIBdecl}{\relax}
\BIBdecl

\bibitem{IMT2021White}
{IMT-2030 (6G) Promotion Group}, ``{White paper on 6G vision and candidate
  technologies},''
  \url{http://www.caict.ac.cn/english/news/202106/t20210608_378637.html}, 2021.

\bibitem{liu2020joint}
F.~Liu, C.~Masouros, A.~Petropulu, H.~Griffiths, and L.~Hanzo, ``Joint radar
  and communication design: {A}pplications, state-of-the-art, and the road
  ahead,'' \emph{IEEE Trans. Commun.}, vol.~68, no.~6, pp. 3834--3862, Jun
  2020.

\bibitem{SLiu2025Random}
S.~Lu, F.~Liu, Y.~Xiong, Z.~Du, Y.~Cui, S.~Li, W.~Yuan, J.~Yang, and S.~Jin,
  ``Sensing with random communication signals,'' \emph{IEEE Network, Early
  Access}, 2025.

\bibitem{sturm2011waveform}
C.~Sturm and W.~Wiesbeck, ``Waveform design and signal processing aspects for
  fusion of wireless communications and radar sensing,'' \emph{Proc. IEEE},
  vol.~99, no.~7, pp. 1236--1259, Jul.

\bibitem{zeng2020joint}
Y.~Zeng, Y.~Ma, and S.~Sun, ``Joint radar-communication with cyclic prefixed
  single carrier waveforms,'' \emph{IEEE Trans. Veh. Technol.}, vol.~69, no.~4,
  pp. 4069--4079, Apr 2020.

\bibitem{Raviteja2018Interference}
P.~Raviteja, K.~T. Phan, Y.~Hong, and E.~Viterbo, ``Interference cancellation
  and iterative detection for orthogonal time frequency space modulation,''
  \emph{IEEE Trans. Wireless Commun.}, vol.~17, no.~10, pp. 6501--6515, Oct
  2018.

\bibitem{bemani2023affine}
A.~Bemani, N.~Ksairi, and M.~Kountouris, ``Affine frequency division
  multiplexing for next generation wireless communications,'' \emph{IEEE Trans.
  Wireless Commun.}, vol.~22, no.~11, pp. 8214 -- 8229, Nov 2023.

\bibitem{Gaudio2020On}
L.~Gaudio, M.~Kobayashi, G.~Caire, , and G.~Colavolpe, ``{On the effectiveness
  of OTFS for joint radar parameter estimation and communication},'' \emph{IEEE
  Trans. Wireless Commun.}, vol.~19, no.~9, pp. 5951--5965, Sep 2020.

\bibitem{hadani2017orthogonal}
R.~H. et~al., ``{Orthogonal time frequency space (OTFS) modulation for
  millimeter-wave communications systems},'' in \emph{Proc. IEEE MTT-S Int.
  Microw. Symp.}\hskip 1em plus 0.5em minus 0.4em\relax IEEE, Jun 2017, pp.
  681--683.

\bibitem{gong2024frequency}
J.~Gong, Y.~Yang, Z.~Wang, D.~Wang, X.~Liu, and M.~Peng, ``Frequency-doppler
  division multiplexing modulation--a waveform design for high doppler
  communications,'' \emph{IEEE Commun. Lett.}, 2024.

\bibitem{luo2024afdm}
Q.~Luo, P.~Xiao, Z.~Liu, Z.~Wan, N.~Thomos, Z.~Gao, and Z.~He, ``{AFDM-SCMA}: A
  promising waveform for massive connectivity over high mobility channels,''
  \emph{IEEE Trans. on Wirel. Commun.}, 2024.

\bibitem{YTao2025IdxAFDM}
Y.~Tao, M.~Wen, Y.~Ge, J.~Li, E.~Basar, and N.~Al-Dhahir, ``Affine frequency
  division multiplexing with index modulation: Full diversity condition,
  performance analysis, and low-complexity detection,'' \emph{IEEE J. Sel.
  Areas Commun., Early Access}, 2025.

\bibitem{Hyin2024MIMOAFDM}
H.~Yin, X.~Wei, Y.~Tang, and K.~Yang, ``Diagonally reconstructed channel
  estimation for {MIMO-AFDM} with inter-doppler interference in doubly
  selective channels,'' \emph{IEEE Trans. Wireless Commun.}, vol.~23, no.~10,
  pp. 14\,066--14\,079, Oct 2024.

\bibitem{ni2022afdm}
Y.~Ni, Z.~Wang, P.~Yuan, and Q.~Huang, ``An {AFDM}-based integrated sensing and
  communications,'' in \emph{Proc. 2022 Int. Symposium on Wireless Commun.
  Syst. (ISWCS)}.\hskip 1em plus 0.5em minus 0.4em\relax IEEE, 2022, pp. 1--6.

\bibitem{bemani2024integrated}
A.~Bemani, N.~Ksairi, and M.~Kountouris, ``Integrated sensing and
  communications with affine frequency division multiplexing,'' \emph{IEEE
  Wirel. Commun. Lett.}, 2024.

\bibitem{yin2022design}
H.~Yin, X.~Wei, Y.~Tang, and K.~Yang, ``{Design and performance analysis of
  AFDM with multiple antennas in doubly selective channels},'' \emph{arXiv
  preprint arXiv:2206.12822}, 2022.

\end{thebibliography}

\end{document}